\documentclass[a4paper]{jpconf}
\usepackage{graphicx}

\usepackage{graphicx}
\usepackage{rotating}
\usepackage{amssymb}
\usepackage{mathptmx}

\begin{document}

\title{Thermal detector model for cryogenic composite detectors for the dark matter experiments CRESST and EURECA}

\author{S~Roth, C~Ciemniak, C~Coppi, F~v~Feilitzsch, A~G\"utlein, C~Isaila, J-C~Lanfranchi, S~Pfister, W~Potzel and W~Westphal}

%\author{C~Ciemniak}, \author{C~Coppi}, \author{F~v~Feilitzsch}, \author{A~G\"utlein}, \author{C~Isaila}, \author{J-C~Lanfranchi}, \author{S~Pfister}, \author{W~Potzel} \and \author{W~Westphal}

\address{Physik-Department E15, Technische Universit\"at M\"unchen, James-Franck-Str., 85748 Garching, Germany}

\ead{sabine.roth@ph.tum.de}

\begin{abstract}
The CRESST (Cryogenic Rare Event Search with Superconducting Thermometers) and the EURECA (European Underground Rare Event Calorimeter Array) experiments are direct dark matter search experiments where cryogenic detectors are used to detect spin-independent, coherent WIMP (Weakly Interacting Massive Particle)-nucleon scattering events by means of the recoil energy. The cryogenic detectors use a massive single crystal as absorber which is equipped with a TES (transition edge sensor) for signal read-out. They are operated at mK-temperatures. In order to enable a mass production of these detectors, as needed for the EURECA experiment, a so-called composite detector design (CDD) that allows decoupling of the TES fabrication from the optimization procedure of the absorber single-crystal was developed and studied. To further investigate, understand and optimize the performance of composite detectors a detailed thermal detector model which takes into account the CDD has been developed.% on the basis of an already existing thermal detector model for cryogenic detectors.% This extended model can be expected to provide an enormous help when tailoring composite detectors to the requirements of various experiments.
\end{abstract}

%\begin{keyword}

%Dark Matter \sep Tungsten TES \sep cryogenic detectors \sep composite detector design \sep thermal detector model

%\PACS 29.40.Mc \sep 63.20.-e \sep 95.35.+d \sep 74.70.Ad

%\end{keyword}

\section{Basics of a cryogenic detector}

The main part of a cryogenic detector, as e.g. used in the CRESST \cite{CRESSTnew} experiment, is given by an absorber single crystal where incident particles (e.g. $\gamma$s, $\beta$s, neutrons or WIMPs \cite{Krauss}) interact by producing a recoil. This energy deposition gives rise to the formation of a phonon population. Read-out of the phonon signal is realized with a so-called TES which is deposited onto the absorber crystal surface. The TES consists of a small ($\sim$mm$^2$) superconducting metal film, acting as thermometer, that is stabilized with the help of a heater in its super-to-normal conducting region where a strong temperature dependence of the resistance can be found. Hence, the basic signal is a resistance change of the TES, that can be read out with a SQUID circuit \cite{SQUID}. A link to the heat bath allows thermal relaxation of the system after an event.

\section{The Composite Detector Design (CDD)}

As for the direct dark matter search a large number of cryogenic detectors (comprising, e.g. for EURECA \cite{Eureca}, a total mass on a ton scale) with very similar properties and the possibility to employ different target materials is needed, the so-called composite detector design (CDD) was developed. It makes it possible to decouple the most critical part of the production process, the deposition of the TES onto the surface of the crystal, from the optimization procedure of the absorber single crystal itself \cite{GNO, ChrisitanI}. The TES is thus fabricated on a small crystal substrate (TES-substrate) which is then coupled to the large absorber crystal by gluing. In this way, the material of the absorber crystal can be chosen freely and does not suffer from the TES production process (e.g. heating cycles). Additionally, using the small crystal substrates allows for the production of several TESs in one step and a simplified pretesting of the TESs' properties before attaching them to the large and expensive absorber crystals.

\section{Thermal detector model}

%In order to further investigate, understand and optimize the performance of composite detectors a detailed thermal detector model including the CDD has been developed on the basis of an already existing thermal detector model for cryogenic detectors \cite{Proebst}.

\subsection{Basic thermal detector model}\label{btdm}

\begin{figure}[!h]
  \begin{center}
   \includegraphics[width=0.85\textwidth, keepaspectratio]{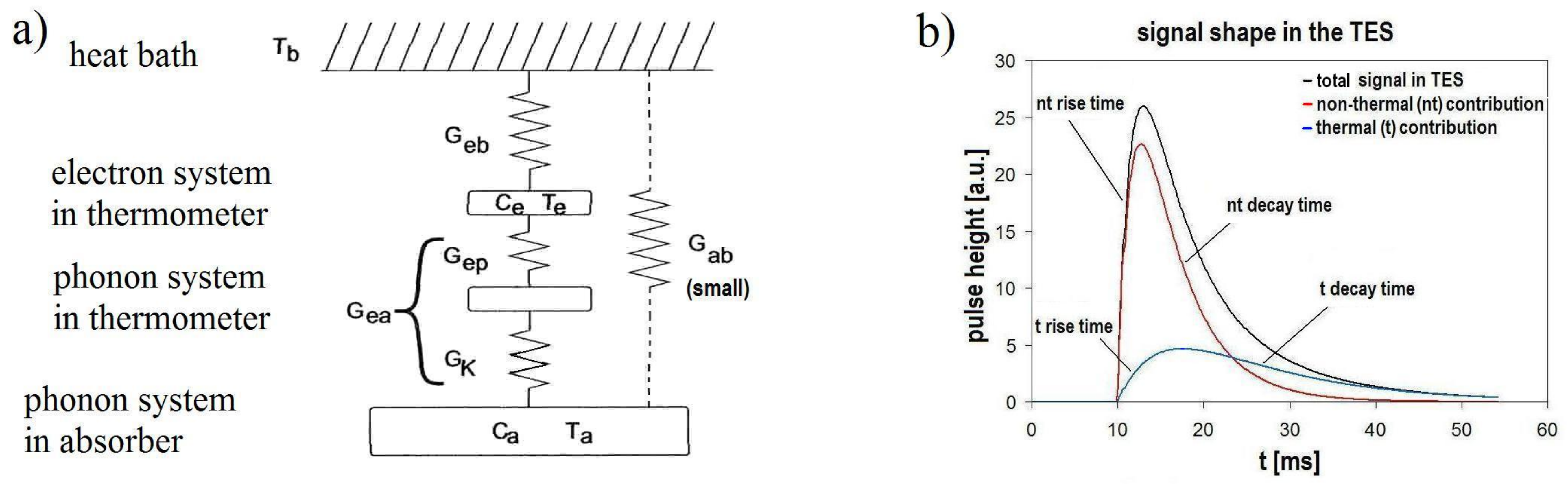}
  \end{center}
  \caption{a) Thermal model of a cryogenic detector with the TES directly deposited on the surface of the absorber \cite{Proebst}. T$_b$, T$_e$ and T$_a$ are the temperatures of the heat bath, the TES electron system and the absorber phonon system, respectively, C$_e$ and C$_a$ are their heat capacities. G$_{ab}$, G$_{eb}$ and G$_{ea}$ are thermal conductances. b) Signal shape in a cryogenic detector with the TES directly deposited on the surface. The respective rise and decay times of the total signal and the thermal (t) and nonthermal (nt) signal components are indicated.}
  \label{thermaldetectormodel}
\end{figure}

The thermal detector model given in \cite{Proebst} describes a classical cryogenic detector in terms of its thermal components, their heat capacitites C$_i$, and the thermal conductances G$_i$ connecting these components (see figure \ref{thermaldetectormodel}a). In this model the resulting pulse shape of the signal (see figure \ref{thermaldetectormodel}b) after an event can be calculated from and explained by these basic detector properties.
%\begin{figure}[!h]
%\includegraphics[width=25pc]{thermaldetectormodel4.pdf}\hspace{2pc}%
%\begin{minipage}[b]{14pc}\caption{\label{thermaldetectormodell}a) Thermal model of a cryogenic detector with the TES (thermometer) directly deposited on the surface of the absorber \cite{Proebst}. T$_b$, T$_e$ and T$_a$ are the temperatures of the heat bath, the TES electron system and the absorber phonon system, respectively, C$_e$ and C$_a$ are their heat capacities. G$_{ab}$, G$_{eb}$ and G$_{ea}$ are thermal conductances. b) Signal shape in a cryogenic detector with the TES directly deposited on the surface with the respective rise and decay times of the total signal and the two (nt and t) signal components.}
%\end{minipage}
%\end{figure}

An event in the absorber crystal immediately creates a nonthermal (nt) phonon population which is not in thermal equilibrium with the system. As can be seen in figures \ref{thermaldetectormodel}a and b, these nt phonons can then either relax via direct thermalisation in the TES, giving rise to a fast (nt) contribution (red) or thermalise in the absorber by inelastic scattering processes at the crystal surfaces leading to a temperature rise of the absorber and therefore delivering a slower, thermal (t) contribution (blue) to the signal. Further details are given in refs. \cite{Proebst}-\cite{Roth}.

\subsection{Thermal detector model for cryogenic composite detectors}

\begin{figure}[h]
\includegraphics[width=29pc]{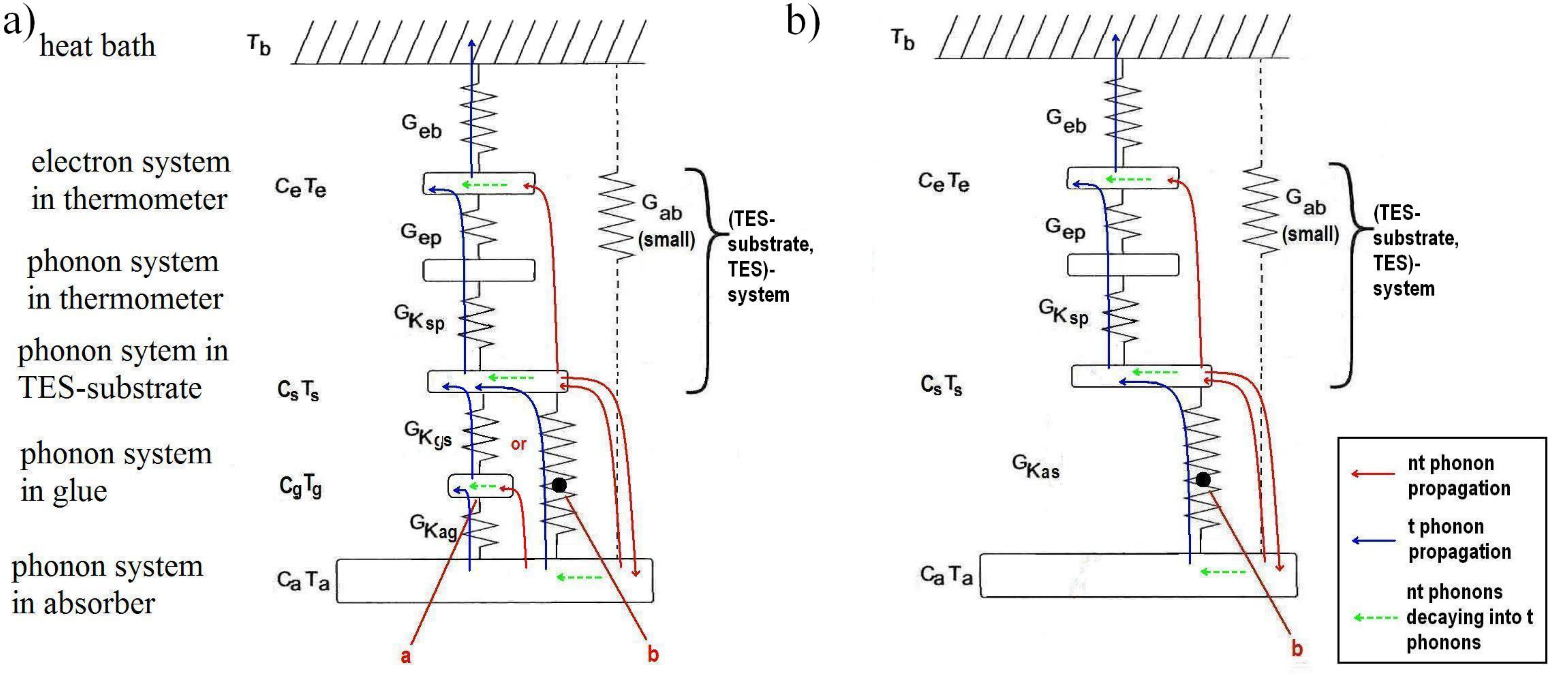}\hspace{1pc}%
\begin{minipage}[b]{7.5pc}\caption{\label{gluetdmphonons}a) Thermal model of a composite detector with two ways of including the glue (\textit{a} as thermal component or \textit{b} as thermal conductance) and the most important phonon propagation possibilities. b) Remaining dominant signal contributions: the glue can be included as thermal conductance only.}
\end{minipage}
\end{figure}

In order to describe a composite detector and its response to an event, i.e. the resulting pulse shape, by a thermal detector model, the glue and the TES-substrate have to be included into the model  and their influence on the evolution of the phonon-population and on the phonon propagation has to be investigated \cite{RothCryoScint, Roth}. In figure \ref{gluetdmphonons}a, a graphical representation describing a composite detector in terms of its thermal components and conductances is given, as well as the most important possibilities for phonon populations to evolve and to propagate (see arrows, red for nt phonons, blue for t phonons). Additionally, two different possibilities of including the glue into the model are shown. It can either be introduced as a thermal component of the system (a), where nt phonons can enter and decay, building up an extra thermal phonon population that can independently contribute to the signal, or it can just be included into the model delivering a thermal conductance between the absorber and the TES-substrate (b).
%\begin{figure}[!h]
%  \begin{center}
%   \includegraphics[width=0.7\textwidth, keepaspectratio]{gluetdmphonons1.pdf}
%  \end{center}
%  \caption{a) Thermal model of a cryogenic composite detector in terms of its thermal components with two different possibilities of including the glue (\textit{a} as thermal component or \textit{b} as contribution to the thermal conductance). b) Most important possibilities of evolving phonon populations and their propagation possibilities.}
%  \label{gluetdmphonons}
%\end{figure}

\noindent The basic concept used for the analysis of this model is to consider the upper part of a composite detector, the (TES-substrate, TES)-system (see figure \ref{gluetdmphonons}a) as an individual cryogenic detector that can be described by the basic thermal detector model, referred to in section \ref{btdm}. However, now the energy input into the system is no longer delivered by an event in its "absorber", the TES-substrate, but is given by nt and t phonons, of the absorber or the glue, entering the TES-substrate via the glue-TES-substrate interface. Thus the shape of the signal in the TES can be obtained by a convolution of the power input from the absorber or the glue with the response of the (TES-substrate, TES)-system \cite{Roth}.

\section{Results}

For each of the phonon populations and propagations indicated in figure \ref{gluetdmphonons}a, the expected signal contribution (pulse rise and decay times) was calculated and then compared to the pulse shapes obtained experimentally for three different realized composite detectors (for details see \cite{RothCryoScint, Roth}). The dominant signal contributions in a composite detector and the underlying mechanisms could be identified successfully (see figure  \ref{gluetdmphonons}b). The main conclusions have been, that the glue is basically transparent for nt phonons and that it has to be included into the model only as a thermal conductance. Since no dominant contribution from nt phonons decaying in the glue could be detected, the glue cannot be contributing a noticeable thermal component to the system. Furthermore, only two different phonon populations were found to dominantly contribute to the signal in the TES: \textit{a)} nt phonons from the absorber can propagate through the glue into the TES-substrate where they can either be directly absorbed in the TES or decay into t phonons and then be absorbed in the TES and \textit{b)} t phonons from the absorber, that already decayed there can contribute to the signal via the thermal conductance delivered by the glue. However, depending on the exact detector design, i.e., the \textit{TES-to-glue-area} ratio, not in every composite detector \textit{both} of these two contributions are significant for the signal in the TES. (For further details see \cite{Roth}.) A discussion concerning the implications on the composite detector design for dark matter experiments like CRESST and EURECA can be found in \cite{RothCryoScint}.

\ack
%\begin{acknowledgements}
This work has been supported by funds of the DFG (SFB 375, Transregio 27: "Neutrinos and Beyond"), the Munich Cluster of Excellence ("Origin and Structure of the Universe"), the EU network for Applied Cryogenic Detectors (HPRN-CT2002-00322), and the \-Maier-Leibnitz-Laboratorium (Garching).
%\end{acknowledgements}

%\bibliography{Neutrbib}

\end{document}